\documentclass[aps,preprint,prd,nofootinbib]{revtex4}

\usepackage{graphicx}
\usepackage{psfrag}
\usepackage{epsfig}

\def\bfej{\mbox{\boldmath$\epsilon$}}
\def\ej{\mbox{$\epsilon$}}

\begin{document}

\title{Dominant Spin-Flip Effects for the Hadronic Produced $J/\psi$ Polarization at TEVATRON}
\author{Xing-Gang Wu \footnote{email:wuxg@itp.ac.cn} and Zhen-Yun Fang}
\address{Department of Physics, Chongqing University, Chongqing 400044,
P.R. China}

\begin{abstract}
Dominant spin-flip effects for the direct and prompt $J/\psi$
polarizations at TEVATRON run II with collision energy $1.96$ TeV
and rapidity cut $|y^{J/\psi}|<0.6$, have been systematically
studied, especially, the spin-flip effect for the transition of
$(c\bar{c})_8[^3S_1]$ into $J/\psi$ has been carefully discussed. It
is found that the spin-flip effect shall always dilute the $J/\psi$
polarization, and with a suitable choice of the parameters $a_{0,1}$
and $c_{0,1,2}$, the $J/\psi$ polarization puzzle can be solved to a
certain degree. At large transverse momentum $p_t$, $\alpha$ for the
prompt $J/\psi$ is reduced by $\sim50\%$ for $f_0 = v^2$ and by
$\sim80\%$ for $f_0=1$. We also study the indirect $J/\psi$
polarization from the $b$-decays, which however is slightly affected
by the same spin-flip effect and then shall provide a
better platform to determine the color-octet matrix elements. \\

\noindent {\bf PACS numbers:} 12.38.Bx, 12.39.Jh, 14.40.Lb

\end{abstract}

\maketitle

\section{Introduction}

Within the non-relativistic QCD (NRQCD) framework \cite{NRQCD}, the
hadronic production of $J/\psi$ is dominated by the gluon
fragmentation in which a gluon fragments into a color-octet state
$(c\bar{c})_8[^3S_1]$. And if the spin-symmetry hold for charm
quarks, as is usually adopted in the literature, then the formed
$J/\psi$ shall always show large transverse polarization at
sufficiently large $p_t$. But this prediction contradicts with the
measured at TEVATRON \cite{oldcdf,expcdfII}. This is the well-known
$J/\psi$ polarization puzzle. Recently, the next-to-leading order
(NLO) analysis of the $J/\psi$ polarization have been done by
Refs.\cite{wang1,wang2}, with $(c\bar{c})$ pair in $J/\psi$ Fock
expansion being in color-singlet state $[^3S_{1}]_1$, color-octets
$[^3S_{1}]_8$ and $[^1S_{0}]_8$ respectively. It is found that even
with those NLO corrections, the $J/\psi$ polarization puzzle still
can not be solved. This implies that by doing the NLO calculations,
one can make the perturbative QCD results more convergent on
$\alpha_s$ and then more reliable, but one can not change the fact
that the produced $J/\psi$ is largely transversely polarized.

There are many suggestions to solve such polarization puzzle, e.g.
the effects of the initial gluon off-shellness may provide a
possible explanation for such $J/\psi$ spin alignment in $p\bar{p}$
collision \cite{ktf}, which depends heavily on the unintegrated
parton distribution of the initial gluon(s). While the spin-flip
interaction provides another effective way to solve the $J/\psi$
polarization puzzle as suggested by Ref.\cite{mwl}. By taking the
spin-flip interactions into account, it has been argued in
Ref.\cite{mwl} that the direct $J/\psi$ and $\psi'$ polarizations
can be diluted to a certain degree. And an analysis for the prompt
$\psi'$ production with proper ranges for the newly introduced
parameters $a_{0,1}$ and $c_{1,2}$ has been presented there. In the
present paper, we shall make a systematical discussion on how the
spin-flip interaction affects the polarization of the direct, the
prompt and the indirect produced $J/\psi$ respectively, which is
much more involved than that of $\psi'$, since we need to consider
contributions from the higher charmonium states accordingly. And
then we shall make a comparison with the newly obtained prompt
$J/\psi$ data by TEVATRON CDF collaboration \cite{expcdfII}.

Assuming the produced $J/\psi$ is measured with the momentum $p$, we
define the rest frame of $J/\psi$ by a Lorentz boost from its moving
frame. The produced $J/\psi$ is polarized with the polarization
vector $\bfej^*$ in the rest frame. Within the framework of NRQCD
factorization \cite{NRQCD}, after decomposing Dirac- and color-
indices, the contribution from the channel through the color-octet
$(c\bar{c})_8[^3S_1]$ to the differential cross section can be
generally written as:
\begin{equation}
d\sigma[(c\bar{c})_8[^3S_1]]= H_{ij} \cdot
T_{ij}(\bfej,\bfej^*,\hat{\bf p}),
\end{equation}
where $\hat{\bf p}=\mathbf{p}/|\mathbf{p}|$, $H_{ij}$ is the
$3\times 3$ spin density matrix for producing $(c\bar{c})_8[^3S_1]$
and $T_{ij}$ is the spin density matrix for the transition of
$(c\bar{c})_8[^3S_1]$ into a polarized $J/\psi$, which can be
decomposed as \cite{mwl}:
\begin{eqnarray}
T_{ij}(\bfej,\bfej^*, \hat{\bf p}) &=& \delta_{ij} \left (
\bfej\cdot \bfej^* a_0+ \bfej\cdot\hat{\bf p}\cdot
\bfej^*\cdot\hat{\bf p} a_1 \right )+(\ej_i \ej_j^* + \ej_j \ej_i^*)
c_0\nonumber\\
&& +\left [ \left (\ej_i \hat p_j +\ej_j \hat p_i\right
)\bfej^*\cdot\hat{\bf p} +(\bfej\leftrightarrow\bfej^*)\right ] c_1
+ \hat p_i \hat p_j \bfej\cdot \bfej^* c_2 , \label{pol}
\end{eqnarray}
where $a_{0,1}$ and $c_{0,1,2}$ are un-determined, non-perturbative
but universal parameters. By taking the heavy quark spin symmetry,
only $c_0=\langle0|{\cal O}_8^{J/\psi}(^3S_1)|0\rangle/6$ is
non-zero, which implies that $J/\psi$ will have the same spin as the
color-octet $(c\bar{c})$ pair. In the present, we shall make a
detailed discussion on the direct $J/\psi$ production through
$p\bar{p}\to J/\psi[n]+X$, where $n$ stands for the intermediate
$(c\bar{c})$-charmonium states up to $v^4$ corrections, i.e.
$n=(^3S_{1})_{1}$, $(^3S_{1})_{8}$, $(^1S_{0})_{8}$ or
$(^3P_{J})_{8}$. Further more, the hard subprocess of the hadronic
process is $ab\to J/\psi[n] +X$, where $ab=gg$, $gq$, $g\bar{q}$ and
$q\bar{q}$ respectively. Based on these processes, the relative
importance of the undetermined parameters $a_{0,1}$ and $c_{0,1,2}$
shall be discussed. The prompt $J/\psi$ polarization shall also be
discussed, whose signal includes $J/\psi$ meson that comes from
decays of the higher charmonium states $\chi_{c1}$, $\chi_{c2}$ and
$\psi'$. Further more, we shall take the same spin-flip effect to
study the indirect $J/\psi$ production from $b$-decays, i.e. $b\to
J/\psi +X$.

The paper is organized as follows. In Sec. II, we present the
calculation technology for the $J/\psi$ production. Numerical
results and discussions are presented in Sec. III, where the results
for the direct, the prompt and the indirect $J/\psi$ polarization
shall be presented. The final section is reserved for a summary.

\section{Calculation technology}

We adopt the same calculation technology as pointed out in
Ref.\cite{mwl} to calculate the direct $J/\psi$ production, and for
self-consistency, we present the calculation procedure in more
detail.

It is more convenient to transform the spin density matrix $T_{ij}$
into a covariant form. For such purpose, we introduce a Lorentz
boost matrix $L^{\mu}_{i}$, whose components can be written as
\cite{chen}
\begin{equation}
L^0_{i}=\frac{p_i}{M} ,\;\;\; L^j_{i}=\delta_{ij}- \frac{p_i p_j}
{\mathbf{p}^2}+ \left(\frac{p_i p_j} {\mathbf{p}^2}\right)
\frac{E_{p}}{M}.
\end{equation}
With the help of such boost matrix, one can transform a purely
space-like four-vector, such as $\ej=(0,\bfej)$, from the rest frame
of $J/\psi$ where the components of $p$ are $(M,\mathbf{0})$ to the
frame in which its components are $p_{\mu}=(E_p,\mathbf{p})$,
$E_{p}=\sqrt{M^2+\mathbf{p}^2}$. By applying $(L^\mu_i)(L^\nu_j)$ to
both sides of Eq.(\ref{pol}), and noting the fact that
\begin{equation}
g_{\mu\nu}L^{\mu}_{i}L^{\nu}_{j}=-\delta_{ij},\;\;\;
L^{\mu}_{i}L^{\nu}_{i}=-g_{\mu\nu}+\frac{p_{\mu} p_{\nu}}{p^2},
\end{equation}
we obtain
\begin{eqnarray}\label{newpol}
T_{\mu\nu}(\ej,\ej^*, \hat{p}) &=&
\left[-g_{\mu\nu}+\frac{p_{\mu}p_{\nu}}{M^2}\right]\left[\ej\cdot
\ej^* (-a_0) + \ej\cdot\hat{p}\cdot \ej^*\cdot\hat{p} a_1 \right]
+\left[\ej_\mu \ej^*_\nu + \ej_\nu \ej_\mu^*\right] c_0\nonumber\\
&& +\left [ \left(\ej_\mu \hat p_\nu +\ej_\nu \hat p_\mu\right )
\ej^*\cdot\hat{p} +(\ej\leftrightarrow\ej^*)\right ] (-c_1) +
\left(\hat p_\mu \hat p_\nu \ej\cdot \ej^* \right) (-c_2) +\cdots,
\end{eqnarray}
where in the $J/\psi$ rest frame,
$\ej^{\mu}=L^{\mu}_{i}\bfej_{i}=(0,\bfej)$ and
$\hat{p}_{\mu}=L^{\mu}_{i}\hat{\bf p}_{i}=(0,\hat{\bf p})$. Further
more, for a particular polarization state $\lambda=(0,\pm1)$,
$T_{\mu\nu}(\ej,\ej^*, \hat{p})$ changes to
$T_{\mu\nu}(\ej(\lambda),\ej^*(\lambda), \hat{p})$.

A convenient measure of $J/\psi$ polarization is the variable
$\alpha=(1-3\xi)/(1+\xi)$, where $\xi=\sigma_L/(\sigma_{L}+
\sigma_{T})$ , $\sigma_{L}$ and $\sigma_{T}$ stand for the
longitudinal and the transverse components of the direct hadronic
cross section respectively. As for the indirect $J/\psi$ production
from $b$ decays, $\xi=\Gamma_L/(\Gamma_{L}+\Gamma_{T})$ with
$\Gamma_{L}$ and $\Gamma_{T}$ stand for the longitudinal and the
transverse components of the indirect decay width respectively.

With the help of Eq.(\ref{newpol}), we can write the total
differential cross section for the direct $J/\psi$ production
through the process $p\bar{p}\to J/\psi^{\lambda}[n]+X$ as:
\begin{equation}\label{totalcs}
d\sigma_{\lambda}(p\bar{p} \to J/\psi^{\lambda}[n] X) = \sum_{ab}
\int dx_a dx_b f_{a/p} (x_a) f_{b/\bar{p}}(x_b)
d\hat\sigma_{\mu\nu}[n, ab] T_{\mu\nu}(\ej(\lambda),\ej^*(\lambda),
\hat{p}),
\end{equation}
where the differential cross sections for the hard subprocesses,
$a(k_1)+b(k_2) \to J/\psi[n](p) +X$, can be written in the following
factorization form,
\begin{eqnarray}\label{totalcs0}
\frac{ d \hat \sigma_{\mu\nu} [n,ab]}{d t} & = & A_{ab}[n]g_{\mu\nu}
+ B_{ab}[n] k_{1 \mu} k_{1 \nu} + C_{ab}[n] k_{2 \mu} k_{2 \nu}
+D_{ab}[n]\frac{1}{2}\left[k_{1 \mu} k_{2 \nu} +k_{2 \mu} k_{1 \nu}
\right ].
\end{eqnarray}
$n$ stands for the $(c\bar{c})$-charmonium state up to $v^4$
corrections, i.e. $n=(^3S_{1})_{1}$, $(^3S_{1})_{8}$,
$(^1S_{0})_{8}$ and $(^3P_{J})_{8}$ respectively. $k_1$ and $k_2$
are the momenta of the initial partons and $ab=gg$, $gq$,
$g\bar{q}$, $q\bar{q}$. The coefficients $A_{ab}[n]$, $B_{ab}[n]$,
$C_{ab}[n]$ and $D_{ab}[n]$ can be read from Refs.\cite{AKL,beneke1}
\footnote{We have calculated all these channels and found a good
agreement with those in Refs.\cite{AKL,beneke1}.}. To calculate the
longitudinal cross section, we adopt the covariant form of the
$J/\psi$ longitudinal polarization vector,
$\ej_{L}(p)_{\mu}=\frac{p\cdot Q} {\sqrt{(p\cdot Q)^2- M^2 Q^2}}
\left(\frac{p_{\mu}}{M}-\frac{M}{p\cdot Q}Q_{\mu}\right)$, where
$Q=p_{p}+p_{\bar{p}}$ is the sum of the initial hadron momenta. In
Refs.\cite{beneke2,braaten}, the fragmentation effect has also been
resummed to the leading logarithms $[\alpha_s\ln p_t^2/(2m_c)^2]^n$
accuracy for the production of the color-octet $^3S_1$ state with
the help of the Altarelli-Parisi evolution equation. By taking the
fragmentation effect into account, the value of $\alpha$ shall be
further suppressed \cite{beneke2}. In this paper, we shall
concentrate our attention on the spin-flip interactions and will not
take this effect into consideration.

Further more, the hadronic differential cross section can be
simplified so as to obtain $J/\psi$ $p_t$ distribution:
\begin{equation}\label{ptcs0}
E_p\frac{d^3\sigma}{d^3p}(p\bar{p} \to J/\psi[n] X)=\sum_{ab}\int
dx_a dx_b f_{a/p}(x_a) f_{b/\bar{p}}(x_b) \times
\frac{\hat{s}}{\pi}\frac{d \hat{\sigma}[n,ab]}
{d\hat{t}}~\delta(\hat{s}+\hat{t}+\hat{u}-M^2) ,
\end{equation}
where $\hat{\sigma}[n,ab]=\hat\sigma_{\mu\nu}[n, ab]
T_{\mu\nu}(\ej,\ej^*, \hat{p})$, and for a particular polarization
state of $J/\psi$,
\begin{eqnarray}
\frac{d\sigma_{\lambda}}{dp_t}(p\bar{p}\rightarrow
J/\psi[n]^{\lambda}X) &=& \sum_{ab} \int dy dx_a
f_{a/p}(x_a)~f_{b/\bar{p}} (x_b^{\prime})\frac{2 p_t}{x_a
x^{\prime}_b (x_a -\frac{M_T}{\sqrt{S}} e^{y})}\nonumber\\
&& \times\frac{d\hat\sigma_{\mu\nu}[n, ab]}{dt}
T_{\mu\nu}(\ej(\lambda),\ej^*(\lambda), \hat{p}),\label{ptcs}
\end{eqnarray}
where
\begin{equation}
x_b^{\prime}=\frac{1}{\sqrt{S}} \frac{x_a \sqrt{S}M_T
e^{-y}-M^2}{x_a \sqrt{S}-M_T e^{y}},
\end{equation}
with $M_T^2=\sqrt{p_t^2+M^2}$, $S=(p_{p}+p_{\bar{p}})^2$ and $y$
stands for the rapidity of $J/\psi$.

Secondly, we present the formulae for the prompt $J/\psi$
polarization. Theoretical predictions of the polarization of prompt
$J/\psi$ are complicated by the fact that the prompt signal includes
$J/\psi$ mesons that also come from decays of the higher charmonium
states $\chi_{cJ}$ ($J=0,1,2$), and $\psi'$. The unpolarized
differential cross-section ${d\sigma^{{\rm prompt}\;
J/\psi}_{tot}}/{dp_t}$ is simply obtained by adding the unpolarized
cross-sections of the various direct-production processes multiplied
with the appropriate branching fractions. While the longitudinal
differential cross-section ${d\sigma^{{\rm prompt}\;
J/\psi}_L}/{dp_t}$ is much more involved, which equals
\begin{equation}
\frac{d\sigma^{{\rm prompt}\; J/\psi}_L}{dp_t}=\frac{d\sigma^{{\rm
direct}\; J/\psi}_L}{dp_t} +\frac{d\sigma^{\chi_{cJ}}_L}{dp_t}
+\frac{d\sigma^{\psi'}_L}{dp_t}
+\frac{d\sigma^{\psi'\to\chi_{cJ}}_L}{dp_t} ,
\end{equation}
where following the discussion in Refs.\cite{braaten,ppol}, we have
\begin{eqnarray}
\sigma^{{\rm direct}\; J/\psi}_L &=&\sum_n\hat\sigma_L(n)\langle
{\cal O}^{J/\psi}(n)\rangle ,\\
\sigma^{\chi_{cJ}}_L &=& \left[\frac{\hat\sigma_{L}(^{3}P^{(1)}_0)}
{3}\langle {\cal O}^{\chi_{c0}}_{1}(^{3}P_0)\rangle
+\frac{\hat\sigma_{L}(^{3}S^{(8)}_1)} {3}\langle {\cal
O}^{\chi_{c0}}_{8}(^{3}S_1)\rangle\right]{\cal B}(\chi_{c0}\to
J/\psi+\gamma)\nonumber\\
&+& \left\{\frac{\hat\sigma_{T}(^{3}P^{(1)}_1)}{2}\langle {\cal
O}^{\chi_{c1}}_{1}(^{3}P_0)\rangle
+\left[\frac{\hat\sigma_{L}(^{3}S^{(8)}_1)}{2}+
\frac{\hat\sigma_{T}(^{3}S^{(8)}_1)}{4}\right]\langle {\cal
O}^{\chi_{c1}}_{8}(^{3}S_1)\rangle\right\}\times\nonumber\\
&&{\cal B}(\chi_{c1}\to J/\psi+\gamma)\nonumber\\
&+&\left\{\left[\frac{2\hat\sigma_{L}(^{3}P^{(1)}_2)}{3}
+\frac{\hat\sigma_{T}(^{3}P^{(1)}_2)}{2}\right]\langle {\cal
O}^{\chi_{c2}}_{1}(^{3}P_0)\rangle
+\left[\frac{17\hat\sigma_{L}(^{3}S^{(8)}_1)}{30}+\right.\right.\nonumber\\
&& \left.\left.\frac{13\hat\sigma_{T}(^{3}S^{(8)}_1)}{60}\right]
\langle{\cal O}^{\chi_{c2}}_{8}(^{3}S_1) \rangle\right\} \times
{\cal B}(\chi_{c2}\to J/\psi+\gamma) \label{br1}\\
\sigma^{\psi'}_L &=&\sigma^{{\rm direct}\; \psi'}_L {\cal
B}(\psi'\to J/\psi+X) \label{br2}
\end{eqnarray}
and
\begin{eqnarray}
\sigma^{\psi'\to\chi_{cJ}}_L &=& \frac{1}{3}\sigma^{{\rm direct}\;
\psi'}_L {\cal B}(\psi'\to \chi_{c0}+\gamma) {\cal B}(\chi_{c0}\to
J/\psi+\gamma)\nonumber\\
&+& \left[\frac{1}{2}\sigma^{{\rm direct}\; \psi'}_L+
\frac{1}{4}\sigma^{{\rm direct}\; \psi'}_T\right]{\cal B}
(\psi'\to \chi_{c1}+\gamma){\cal B}(\chi_{c1}\to J/\psi+\gamma)\nonumber\\
&+&\left[\frac{17}{30}\sigma^{{\rm direct}\;
\psi'}_L+\frac{13}{60}\sigma^{{\rm direct}\; \psi'}_T\right] {\cal
B}(\psi'\to \chi_{c2}+\gamma){\cal B}(\chi_{c2}\to J/\psi+\gamma) .
\label{br3}
\end{eqnarray}
$\hat\sigma_L(n)$ and $\hat\sigma_T(n)$ stand for the longitudinal
and transverse cross sections without the matrix elements
accordingly, and the summation in the first equation is over
$n=(^{3}S_{1})_1$, $(^{3}S_1)_{8}$, $(^{1}S_0)_8$ and $(^{3}P_J)_8$.
The direct $\psi'$ production $\sigma^{{\rm direct}\; \psi'}_L
=\sum_n\hat \sigma_L(n)\langle {\cal O}^{\psi'}(n)\rangle$ can be
obtained from that of $J/\psi$ by changing the $J/\psi$ matrix
elements to $\psi'$ matrix elements. The cross-section for
$n=(c\bar{c})_1(^3P_J)$ is much more involved and we put some
necessary formulae in the APPENDIX A.

Finally, by taking the spin-flip effect into consideration, we
recalculate the $J/\psi$ production from the $b$-decay process,
$b\to J/\psi[n]+X$. Following the same procedure of
Ref.\cite{btopsi}, the total unpolarized cross-section can be
written as
\begin{eqnarray}
\Gamma_{\rm tot}(b \rightarrow J/\psi + X) & = &
\frac{G_F^2}{144\pi}\Big| V_{cb} \Big|^2 m_c
m_b^3\left(1-\frac{4m_c^2}{m_b^2}\right)^2\times
\left[a\left(1+\frac{8m_c^2}{m_b^2}\right)+b\right],
\label{bpsi-unpol}
\end{eqnarray}
where
\begin{eqnarray}
a & = & \Bigg(3(C_+ + C_-)^2\times \left[\frac{[2c_0+3a_0+a_1]+
\frac{m_b^2}{m_b^2+8m_c^2}[4c_1+3c_2]} {2m_c^2} + \frac{\langle{\cal
O}^{J/\psi}_8({^3P_0})\rangle}
{m_c^4}\right]\nonumber\\
&&+ (2C_+ - C_-)^2 \frac{\langle{\cal O}^{J/\psi}_1({^3S_1})\rangle}
{3m_c^2}\Bigg), \label{para-a} \\
b &=& 3(C_+ + C_-)^2 \frac{\langle{\cal
O}^{J/\psi}_8(^1S_0)\rangle}{2m_c^2} . \label{para-b}
\end{eqnarray}
The Wilson coefficients $C_+(m_b)=0.868$ and $C_{-}(m_b)=1.329$
\cite{wilson}. In Eq.(\ref{bpsi-unpol}), upon summing over the light
quarks $s$ and $d$, we have used $|V_{cb}V_{cs}^*|^2 +
|V_{cb}V_{cd}^*|^2 \approx |V_{cb}|^2$. We have taken the initial
$b$-quark to be unpolarized. Here $\langle{\cal
O}^{J/\psi}_n\rangle\equiv\langle 0|{\cal O}^{J/\psi}_n|0\rangle$
are NRQCD $J/\psi$ production matrix elements. If neglecting the
spin-flip effects, we return to the same results derived by
Ref.\cite{btopsi}. As for the longitudinal cross-section from $b$
decays, we obtain
\begin{eqnarray}
\Gamma_{\rm L}(b \rightarrow {J/\psi} + X) & = &
\frac{G_F^2}{864\pi}\; \frac{(m_b^2 - 4 m_c^2)^2}{m_b^3 m_c} \Big|
V_{cb} \Big|^2 \;\; \Bigg\{ 2(2C_+ - C_-)^2 m_b^2 \big\langle{\cal
O}^{J/\psi}_1({^3S_1})
\big\rangle \nonumber\\
& + & 9(m_b^2+8m_c^2)(C_+ + C_-)^2\left(
[a_0+a_1]+\frac{m_b^2}{m_b^2+8m_c^2}[2c_0+4c_1+c_2]\right)\nonumber\\
&+& 3m_b^2 (C_+ + C_-)^2 \big\langle{\cal O}^{J/\psi}_8({^1S_0})
\big\rangle +72m_c^2(C_+ + C_-)^2\frac{\big\langle{\cal
O}^{J/\psi}_8({^3P_0}) \big\rangle}{m_c^2}\Bigg\}. \label{bpsi-long}
\end{eqnarray}
In these equations, we have adopted the relation $\langle{\cal
O}^{J/\psi}_8({^3P_J})\rangle=(2J+1)\langle{\cal
O}^{J/\psi}_8({^3P_0})\rangle$ \cite{NRQCD}.

\section{Numerical results and discussions}

\begin{table}
\begin{tabular}{|c|c|c|c|c|c|c|c|c|c|}
\hline $\langle O^{J/\psi}_1(^3S_1)\rangle$& $\langle
O^{J/\psi}_8(^3S_1)\rangle$& $M_{3.4}^{J/\psi}$& $\langle
O^{\psi^\prime}_1(^3S_1)\rangle$& $\langle
O^{\psi^\prime}_8(^3S_1)\rangle$& $M_{3.5}^{\psi^\prime}$ & $\langle
O^{\chi_{c0}}_8(^3P_0)\rangle$ & $\langle O^{\chi_{c0}}_8(^3S_1)\rangle$\\
\hline $1.4\pm 0.1$& $3.9\pm 0.7$& $6.6\pm 0.7$& $6.7\pm
0.7$& $3.7\pm0.9$& $0.78\pm 0.36$ & $9.1\pm1.3$ & $1.9\pm0.2$ \\
GeV$^3$& $10^{-3}$GeV$^3$& $10^{-2}$GeV$^3$& $10^{-1}$GeV$^3$&
$10^{-3}$GeV$^3$& $10^{-2}$GeV$^3$ & $10^{-2}$GeV$^5$ & $10^{-3}$GeV$^3$\\
\hline
\end{tabular}
\caption{Adopted NRQCD matrix elements from Ref.\cite{braaten}.}
\label{tab1}
\end{table}

As for numerical calculation, we adopt the matrix elements derived
by Ref.\cite{braaten}, which are shown in TAB.\ref{tab1}. To be
consistent, the parton distribution function is chosen to be CTEQ5L
\cite{cteq5} and the value of $\alpha_s$ is evaluated from the
one-loop formula using the corresponding value in CTEQ5L for
$\Lambda_{QCD}$. $m_c=1.5GeV$ and $M_{J/\psi}=2m_c$. Both the
factorization scale and the renormalization scale are taken to be
the transverse mass of $J/\psi$, i.e.
$\mu_{f}=\mu_{r}=\sqrt{M^2+p_t^2}$. The collision center of mass
(C.M.) energy is $1.96$ TeV. The branching ratio of
$J/\psi\to\mu^{+}\mu^{-}$ is $\beta=(5.93\pm0.06)\times 10^{-2}$
\cite{pdg}. As for the charmonium, we take $v^2=0.30$.

There are two types of power counting rules for the undetermined
parameters $a_{0,1}$ and $c_{0,1,2}$ \cite{mwl,count2}, i.e. the
first type is
\begin{equation}\label{rule1}
\frac{a_0 }{c_0}\sim v^2 \quad  \frac{a_1 }{c_0}=\frac{c_1 }{c_0}=
\frac{c_2 }{c_0}\sim v^3 \,
\end{equation}
and the second type is
\begin{equation}\label{rule2}
\frac{a_0 }{c_0}\sim  \frac{a_1 }{c_0} \sim  \frac{c_1 }{c_0}\sim
\frac{c_2 }{c_0}\sim {\mathcal O} (1).
\end{equation}
Since the value of $v$ is generally not small for the case of
charmonium, those higher $v$-suppressed terms can have a significant
impact on theoretical predictions even for the first type of power
counting rule.

\subsection{relative importance among the different terms in
$T_{\mu\nu}(\ej,\ej^*, \hat{p})$}

This subsection is served to show the relative importance among the
different terms in $T_{\mu\nu}(\ej,\ej^*, \hat{p})$. In order to
show clearly the contributions from each parts of
$T_{\mu\nu}(\ej,\ej^*, \hat{p})$, we rewrite $\frac{d\sigma[ab,n]}
{dp_t}$ as ($n=(^3S_{1})_{8}$),
\begin{equation}
\frac{d\sigma[ab,n]}{dp_t}=\frac{d\sigma^{c_0}[ab,n]} {dp_t}c_0
+\frac{d\sigma^{c_1}[ab,n]}{dp_t}c_1+
\frac{d\sigma^{c2}[ab,n]}{dp_t}c_2+
\frac{d\sigma^{a_0}[ab,n]}{dp_t}a_0+
\frac{d\sigma^{a_1}[ab,n]}{dp_t}a_1 ,
\end{equation}
where $\frac{d\sigma^{c_0}[ab,n]} {dp_t}$ stands for the
$p_t$-distribution for $c_1=c_2=a_0=a_1=0$ with an overall factor
$c_0$ being contracted out, and etc.. To show the relative
importance among different distributions, we define the ratio,
\begin{equation}\label{relativeP}
R^{c_1}[ab,n]=\frac{\frac{d\sigma^{c_1}[ab,n]}{dp_t}}
{\frac{d\sigma^{c_0}[ab,n]} {dp_t}},\;\;
R^{c_2}[ab,n]=\frac{\frac{d\sigma^{c_2}[ab,n]}{dp_t}}
{\frac{d\sigma^{c_0}[ab,n]} {dp_t}},\;\;
R^{a_0}[ab,n]=\frac{\frac{d\sigma^{a_0}[ab,n]}{dp_t}}
{\frac{d\sigma^{c_0}[ab,n]} {dp_t}},\;\;
R^{a_1}[ab,n]=\frac{\frac{d\sigma^{a_1}[ab,n]}{dp_t}}
{\frac{d\sigma^{c_0}[ab,n]} {dp_t}},
\end{equation}
where $ab=gg$, $gq$, $g\bar{q}$ and $q\bar{q}$.

\begin{figure}
\centering
\includegraphics[width=0.47\textwidth]{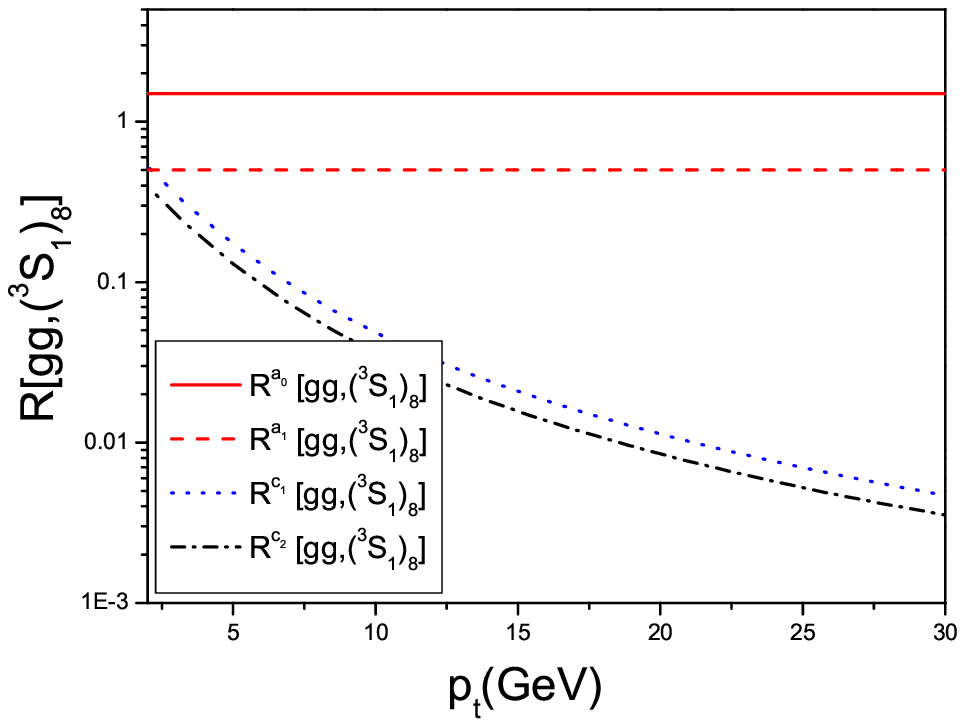}%
\includegraphics[width=0.48\textwidth]{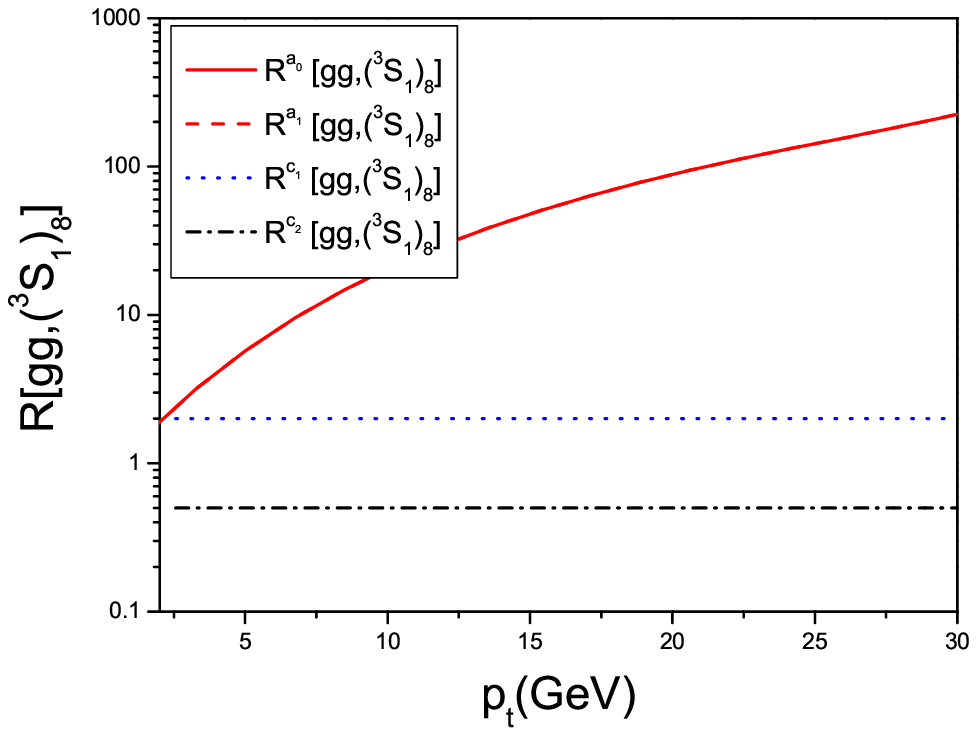}%
\caption{$R[gg,{(^3S_1)_{8}}]$-distributions defined by
Eq.(\ref{relativeP}), where the left diagram is derived by summing
over all the polarizations and the right diagram is only for the
longitudinal polarization of $J/\psi$. For the right diagram, the
cures for $a_0$ and $a_1$ are coincide with each other.} \label{rr}
\end{figure}

The $R$-distributions for the dominant gluon-gluon fusion mechanism
($ab=gg$) are shown in Fig.(\ref{rr}), where the left diagram is
derived by summing over all the polarizations and the right diagram
is only for the longitudinal polarization of $J/\psi$. The left
diagram of Fig.(\ref{rr}) shows that when summing over all the
polarization vectors of $J/\psi$, the weights of $a_0$ and $a_1$
shall always at the same order of that of $c_0$, and more explicitly
$R^{a_0}[gg,(^3S_{1})_{8}]\equiv 3/2$ and
$R^{a_1}[gg,(^3S_{1})_{8}]\equiv 1/2$; while those of $c_1$ and
$c_2$ drop down quickly with the increment of $p_t$ (${\cal O}
(1/p_t^2)$). The right diagram of Fig.(\ref{rr}) shows that for only
the longitudinal part, the weights of $a_0$ and $a_1$ increase
quickly in comparison with that of $c_0$ with the increment of $p_t$
${\cal O}(p_t^2)$; while the weights of $c_1$ and $c_2$ have the
same order of that of $c_0$, or more explicitly
$R^{c_1}[gg,(^3S_{1})_{8}]\equiv 2$ and
$R^{c_2}[gg,(^3S_{1})_{8}]\equiv 1/2$.

\subsection{a simple discussion on the color-octet $^3S_1$
matrix element under the spin-flip interaction}

Summing over the polarizations on both sides of Eq.(\ref{pol}), we
obtain a new matrix element for the color-octet $^3S_1$ state,
\begin{equation}\label{parar1}
\langle 0 \vert O_8^{J/\psi} ( ^3 S_1 ) \vert 0\rangle^\prime = 6
c_0\left [ 1 +\frac{3 a_0}{2 c_0} +  \frac{a_1}{2 c_0}  +
\frac{2c_1}{3 c_0} + \frac{c_2}{2 c_0} \right ] .
\end{equation}
In principle, the value of the new matrix element $\langle0 \vert
{\cal O}_{8}^{J/\psi}(^3S_1)\vert 0\rangle^{\prime}$ defined in
Eq.(\ref{parar1}) is different from that of the usual $\langle0
\vert {\cal O}_{8}^{J/\psi}(^3S_1)\vert 0\rangle$, since it involves
an extra gauge link in the definition and it also takes the
spin-flip interaction into account. By taking into the spin-flip
interaction and the extra gauge links, the matrix element $\langle0
\vert {\cal O}_{8}^{J/\psi}(^3S_1)\vert 0\rangle^{\prime}$ contains
more non-perturbative parameters to be determined, i.e. $a_0$,
$a_1$, $c_0$, $c_1$ and $c_2$. In the following, we shall discuss
the spin-flip effects based on the two counting rules (\ref{rule1})
and (\ref{rule2}), and to predict the polarization, we make the
ansatz that $a_1=c_1=c_2$ and introduce two parameters $f_0=a_0/c_0$
and $f_1=a_1/c_0$, we obtain
\begin{equation}\label{para1}
\langle{\cal O}^{J/\psi}_{8}(^3S_1)\rangle^{\prime}=6c_0
\left(1+\frac{3}{2}f_0 + \frac{5}{3}f_1 \right) .
\end{equation}
$f_1=v f_0$ for the power counting rule (\ref{rule1}), $f_1=f_0$ for
the power counting rule (\ref{rule2}) respectively. By setting
$f_0=f_1=0$, we return to the result without taking into account the
spin-flip interaction, which leads to $c_0=\langle0 \vert {\cal
O}_{8}^{J/\psi}(^3S_1)\vert 0\rangle/6$.

\begin{figure}
\centering
\includegraphics[width=0.460\textwidth]{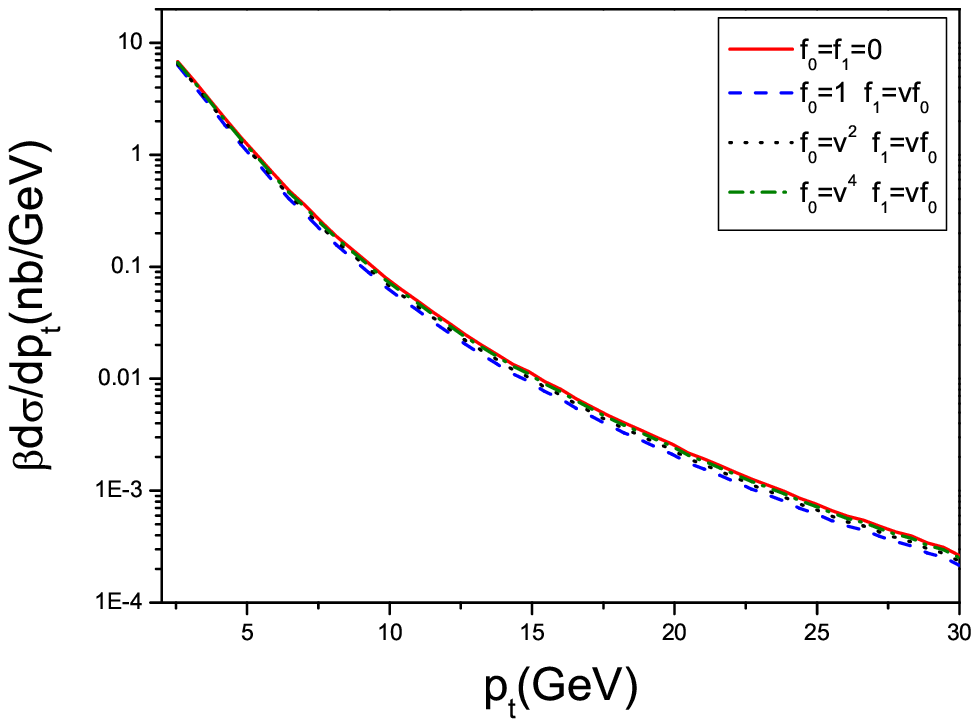}%
\hspace{0.1cm}
\includegraphics[width=0.460\textwidth]{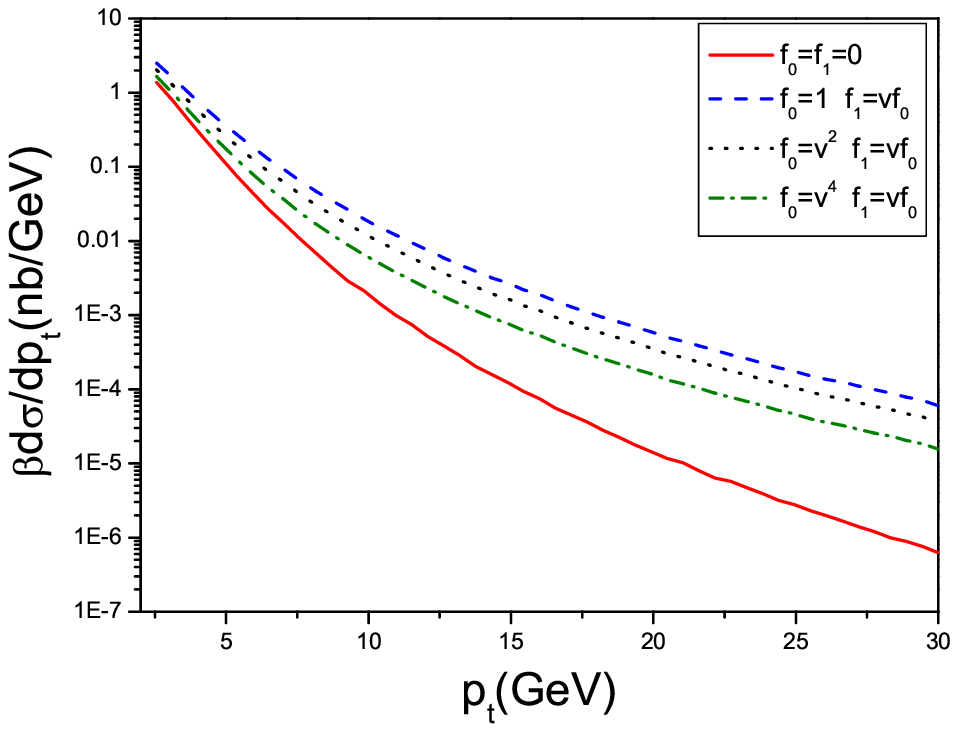}%
\caption{$p_t$-distributions for the direct production of $J/\psi$
through $(c\bar{c})_8[^3 S_1]$ at C.M. energy $\sqrt{s}=1.96$ TeV.
Left diagram is derived by summing over all the polarizations and
the right diagram is only for the longitudinal polarization of
$J/\psi$. $\beta$ stands for the branching ratio
$J/\psi\to\mu^{+}\mu^{-}$. In the calculation, the contributions
from all the considered subprocesses have been summed up, and
rapidity cut $|y^{J/\psi}<0.6|$ is adopted.} \label{totpt1}
\end{figure}

In Fig.(\ref{totpt1}), we show the $p_t$-distributions with
different value of $f_0$ and $f_1=v f_0$ for the direct production
of $J/\psi$ through $(c\bar{c})_{8}[^3S_{1}]$ at TEVATRON run II
($\sqrt{s}=1.96TeV$), where the left diagram is derived by summing
over all the polarizations and the right diagram is only for the
longitudinal polarization of $J/\psi$. In the calculation, the
contributions from all the considered subprocesses (with
$ab=gg,gq,g\bar{q},q\bar{q}$) have been summed up, and the rapidity
cut $|y^{J/\psi}<0.6|$ is adopted. Fig.(\ref{totpt1}) shows that by
considering the spin-symmetry-breaking factors $a_{0,1}$ and
$c_{1,2}$, the $p_t$-distributions derived by summing over all the
polarizations of $J/\psi$ shall not be affected too much in
comparison with the case of not considering the spin flip
interactions (with $f_0=f_1=0$). While for the longitudinal
$p_t$-distributions, the spin-symmetry-breaking factors $a_{0,1}$
and $c_{1,2}$ might be important in the large $p_t$ regions, i.e.
they can raise the longitudinal contributions to a certain degree.
So such spin symmetry breaking corrections can not be ignored safely
as has been done in Ref.\cite{beneke3}, since as shown in the left
diagram of Fig.(\ref{totpt1}), they can change the fraction of the
longitudinal part dramatically. To show this point more clearly, we
present the differential cross-section formulae for $J/\psi$
production channel through $g+g\to J/\psi((^3S_1)_8)+g$ in APPENDIX
B. It can be found that $c_0$ comes into contributions at ${\cal
O}(1/p_t^2)$ for the longitudinal differential cross section. This
is the reason why by taking the spin-symmetry, $a_0=a_1=c_1=c_2=0$,
the longitudinal contributions should be neglected at large $p_t$
regions, i.e. the $J/\psi$ is transverse polarized at large $p_t$
regions. And it is clear that longitudinal differential cross
section will not be suppressed by $1/p_\perp^2$, if one takes
spin-flip interaction into account, i.e., if those coefficients
beside $c_0$, especially $a_0$ and $a_1$, are not zero.

In the above, we have shown that the spin-symmetry-breaking factors
$a_0$, $a_1$, $c_1$ and $c_2$ might be important in the large $p_t$
regions for the production through $(c\bar{c})_{8}[^3S_{1}]$, i.e.
they can raise the longitudinal contributions to a certain degree.
In the large $p_t$ regions, it is the $(c\bar{c})_{8}[^3S_{1}]$
production channel that yields the transverse polarization, while
the $(c\bar{c})_{8}[^1S_{0}]$ and $(c\bar{c})_{8}[^3P_{J}]$ channels
both yield unpolarized quarkonia in this limit \footnote{We have
also calculated the spin-flip effects for $(c\bar{c})_{8}[^1S_{0}]$,
which is quite small in comparison to these of
$(c\bar{c})_{8}[^3S_{1}]$, so we shall not take these spin-flip
effects into consideration. }. So, the large transverse polarization
might be considerably changed by the spin-symmetry-breaking factors
$a_0$, $a_1$, $c_1$ and $c_2$, and then the large discrepancy
between the theoretical prediction and the experimental data might
be compensated.

\subsection{Numerical results of $\alpha$ for direct $J/\psi$ production}

The polarization is predicted with the parameter $\alpha$ as a
function of $p_\perp$, which is defined as:
\begin{equation}
\alpha = \left ( \frac{d \sigma_{tot}}{dp_\perp} -3\frac{d
\sigma_{L}}{dp_\perp} \right ) {\Big /} \left ( \frac{d
\sigma_{tot}}{dp_\perp} +\frac{d \sigma_{L}}{dp_\perp} \right
)=\frac{1-3\xi}{1+\xi},
\end{equation}
where $\xi=\frac{d \sigma_{L}/dp_\perp}{d \sigma_{tot}/dp_\perp}$.
If $\alpha=1$, the produced $\psi$ is transversely polarized. If
$\alpha=-1$ the produced $\psi$ is longitudinally polarized.

After summing over all the above mentioned channels, we represent
the longitudinal polarization fraction $\xi$ for the direct $J/\psi$
production as
\begin{equation}
\xi(f_0,f_1)=\frac{\frac{d\hat\sigma_L^{(^3S_1)_1}}{dp_t}\langle{\cal
O}^{J/\psi}_1(^3S_1)\rangle+\frac{d\hat\sigma_L^{(^3S_1)_8}}{dp_t}\langle{\cal
O}^{J/\psi}_8(^3S_1)\rangle+\left(x\frac{d\hat\sigma_L^{(^1S_0)_8}}{dp_t}
+(1-x)\frac{d\hat\sigma_L^{(^3P_J)_8}}{dp_t}\right)M_r}
{\frac{d\hat\sigma_{tot}^{(^3S_1)_1}}{dp_t}\langle{\cal
O}^{J/\psi}_1(^3S_1)\rangle+\frac{d\hat\sigma_{tot}^{(^3S_1)_8}}{dp_t}
\langle{\cal O}^{J/\psi}_8(^3S_1)\rangle+
\left(x\frac{d\hat\sigma_{tot}^{(^1S_0)_8}}{dp_t}+
(1-x)\frac{d\hat\sigma_{tot}^{(^3P_J)_8}}{dp_t}\right)M_r},
\end{equation}
where $\frac{d\hat\sigma_L^{(^3S_1)_1}}{dp_t}$ stands for the
differential cross section without the matrix element $\langle{\cal
O}^{J/\psi}_1(^3S_1)\rangle$ for the case of $J/\psi$ through
color-singlet $(^3S_1)$-charmonium state, and so on.
$M^{J/\psi}_r=\langle{\cal
O}^{J/\psi}_{8}(^1S_0)\rangle+r\langle{\cal
O}^{J/\psi}_{8}(^3P_0)\rangle/m_c^2$ with $r=3.4$ \cite{braaten},
and the parameter $x=\langle{\cal
O}^{J/\psi}_8(^1S_0)\rangle/M^{J/\psi}_r$, whose center value is
$1/2$ and can be varied between 0 to 1.

\begin{figure}
\centering
\includegraphics[width=0.460\textwidth]{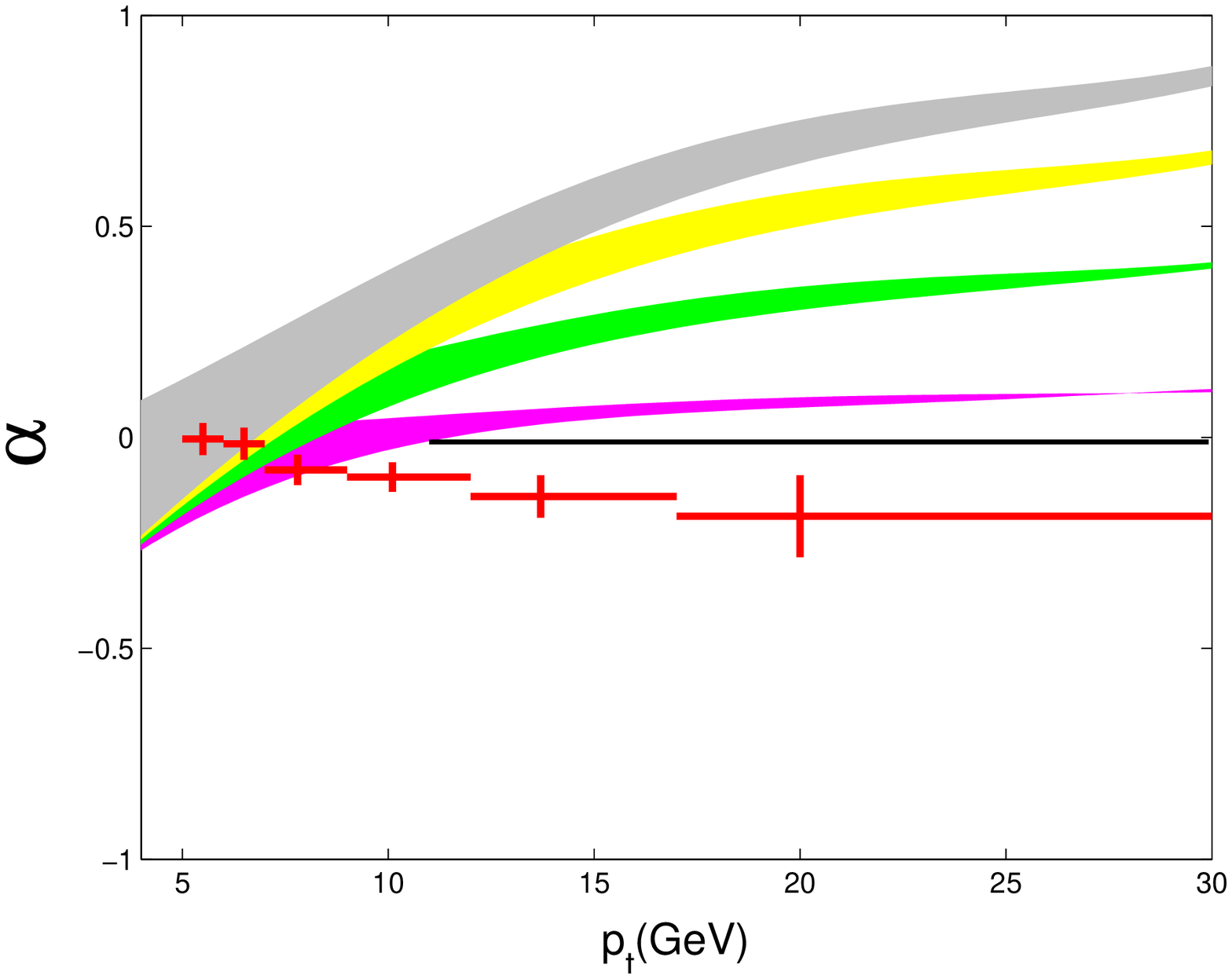}%
\hspace{0.2cm}
\includegraphics[width=0.460\textwidth]{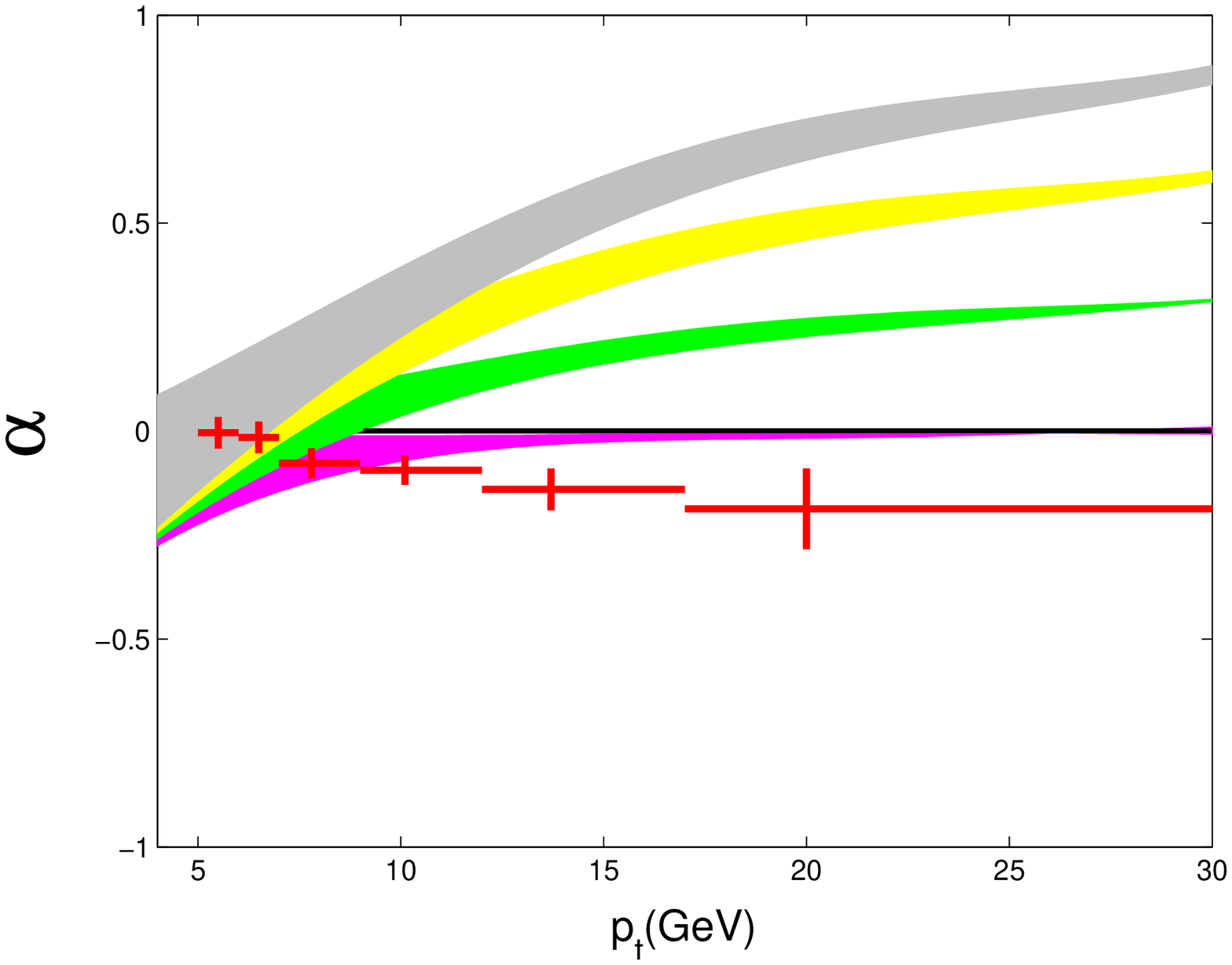}
\caption{$\alpha$ for the direct $J/\psi$ production. Left diagram
is for $f_1=v f_0$ and the right diagram is for $f_1=f_0$. The upper
shaded band is for $f_0=0$, the higher middle yellow band is for
$f_0=v^4$, the lower middle green band is for $f_0=v^2$ and the
lowest band is for $f_0=1$, where the uncertainties of the matrix
elements are also included. The upper edge of each band is for $x=1$
and the lower edge for $x=0$. The rapidity cut $|y^{J/\psi}|<0.6$ is
adopted. The experimental data is from Ref.\cite{expcdfII}.}
\label{dpsi}
\end{figure}

In Fig.(\ref{dpsi}), we present the value of $\alpha$ as a function
of $p_t$ for the direct $J/\psi$ hadronic production, where typical
values for $f_0$ and $f_1$ are adopted and the uncertainties from
the matrix elements are also included. From the figure the produced
$J/\psi$ will be dominantly with transverse polarization at large
$p_\perp$, if one does not take the spin-flip interaction into
account, i.e., $f_0=f_1=0$. Increasing $f_0$ and $f_1$ from $0$ to
$1$, $\alpha$ will be decreased accordingly. And if one takes $f_0$
and $f_1$ at the order of $1$, $\alpha$ shall be close to $0$, which
implies that the $J/\psi$ is unpolarized.

\subsection{Numerical results of $\alpha$ for the prompt $J/\psi$ production}

In calculating $\alpha$ for the prompt $J/\psi$ production, we need
to know the non-perturbative matrix elements $\langle {\cal
O}^{\chi_{cJ}}_{1}(^{3}P_J)\rangle$ and $\langle {\cal
O}^{\chi_{cJ}}_{8}(^{3}S_1)\rangle$. And for their values we adopt
the following relations:
\begin{eqnarray}
\langle {\cal O}^{\chi_{cJ}}_{1}(^{3}P_J)\rangle &=& (2J+1)\langle
{\cal O}^{\chi_{c0}}_{1}(^{3}P_0)\rangle \\
\langle {\cal O}^{\chi_{cJ}}_{8}(^{3}S_1)\rangle &=& (2J+1)\langle
{\cal O}^{\chi_{c0}}_{8}(^{3}S_1)\rangle
\end{eqnarray}
$\langle{\cal O}^{\chi_{c0}}_{1}(^{3}P_0)\rangle =(9.1\pm1.3)\times
10^{-2}$ $GeV^5$ and $\langle {\cal
O}^{\chi_{c0}}_{8}(^{3}S_0)\rangle =(1.9\pm 0.2)\times10^{-3}$
$GeV^3$. As for the relevant branching fractions listed in
Eqs.(\ref{br1},\ref{br2},\ref{br3}), we adopt the values from
Ref.\cite{pdg}, i.e. ${\cal B}(\chi_{c0}\to
J/\psi+\gamma)=(1.30\pm0.11)\%$, ${\cal B}(\chi_{c1}\to
J/\psi+\gamma)=(35.6\pm1.9)\%$, ${\cal B}(\chi_{c2}\to
J/\psi+\gamma)=(20.2\pm1.0)\%$, ${\cal B}(\psi'\to
J/\psi+X)=(56.1\pm0.9)\%$, ${\cal B}(\psi'\to
\chi_{c0}+\gamma)=(9.2\pm0.4)\%$, ${\cal B}(\psi'\to
\chi_{c1}+\gamma)=(8.7\pm0.4)\%$, ${\cal B}(\psi'\to
\chi_{c2}+\gamma)=(2.6\pm0.4)\%$. As for the direct production
$\sigma^{{\rm direct}\; J/\psi}$ and $\sigma^{{\rm direct}\;
\psi'}$, only the spin-flip effects in the channel of
$(c\bar{c})_8[^{3}S_1]$ are sizable. So we shall only consider the
spin-flip effects in the direct production $\sigma^{{\rm direct}\;
J/\psi}$ and $\sigma^{{\rm direct}\; \psi'}$.

\begin{figure}
\centering
\includegraphics[width=0.460\textwidth]{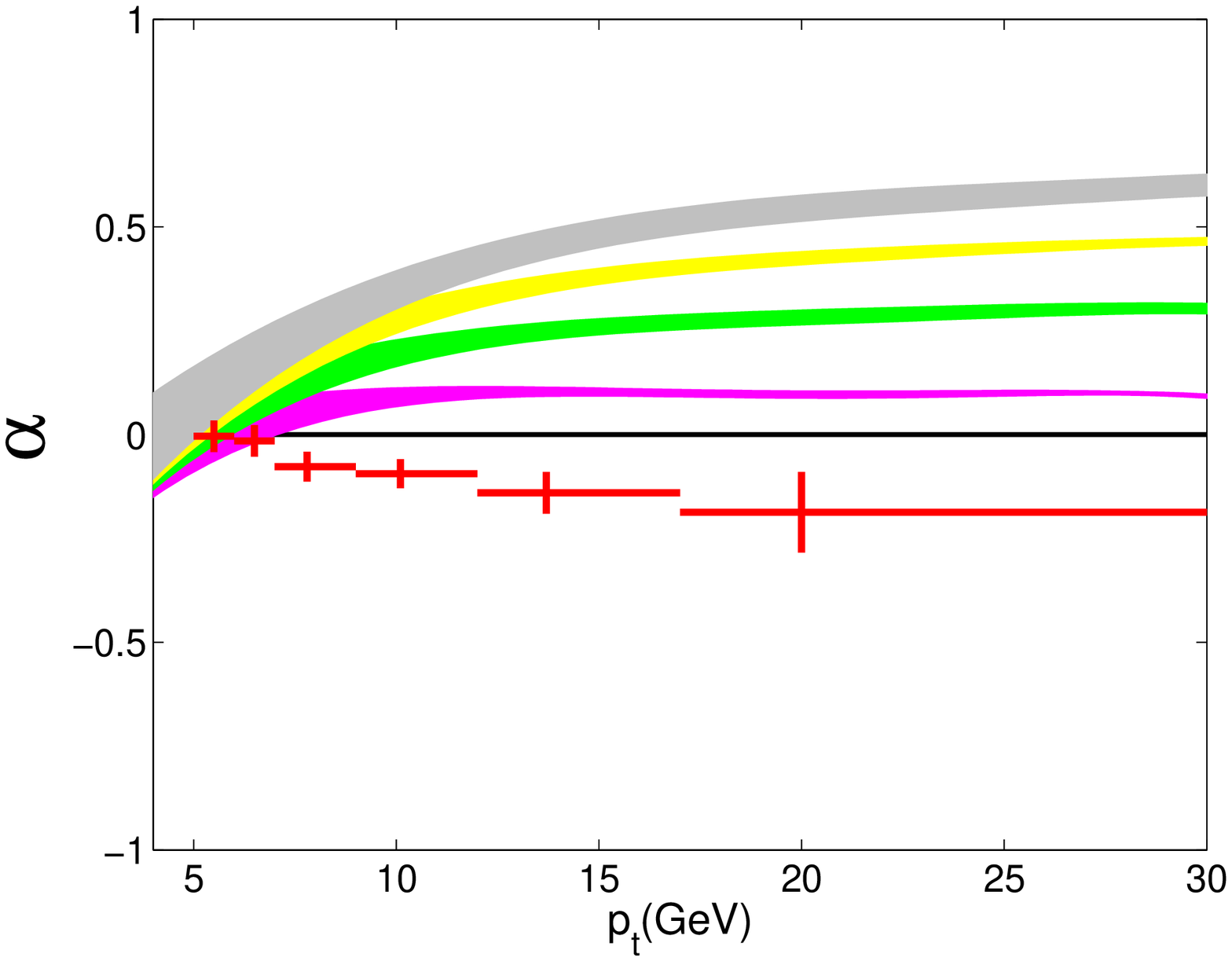}
\includegraphics[width=0.460\textwidth]{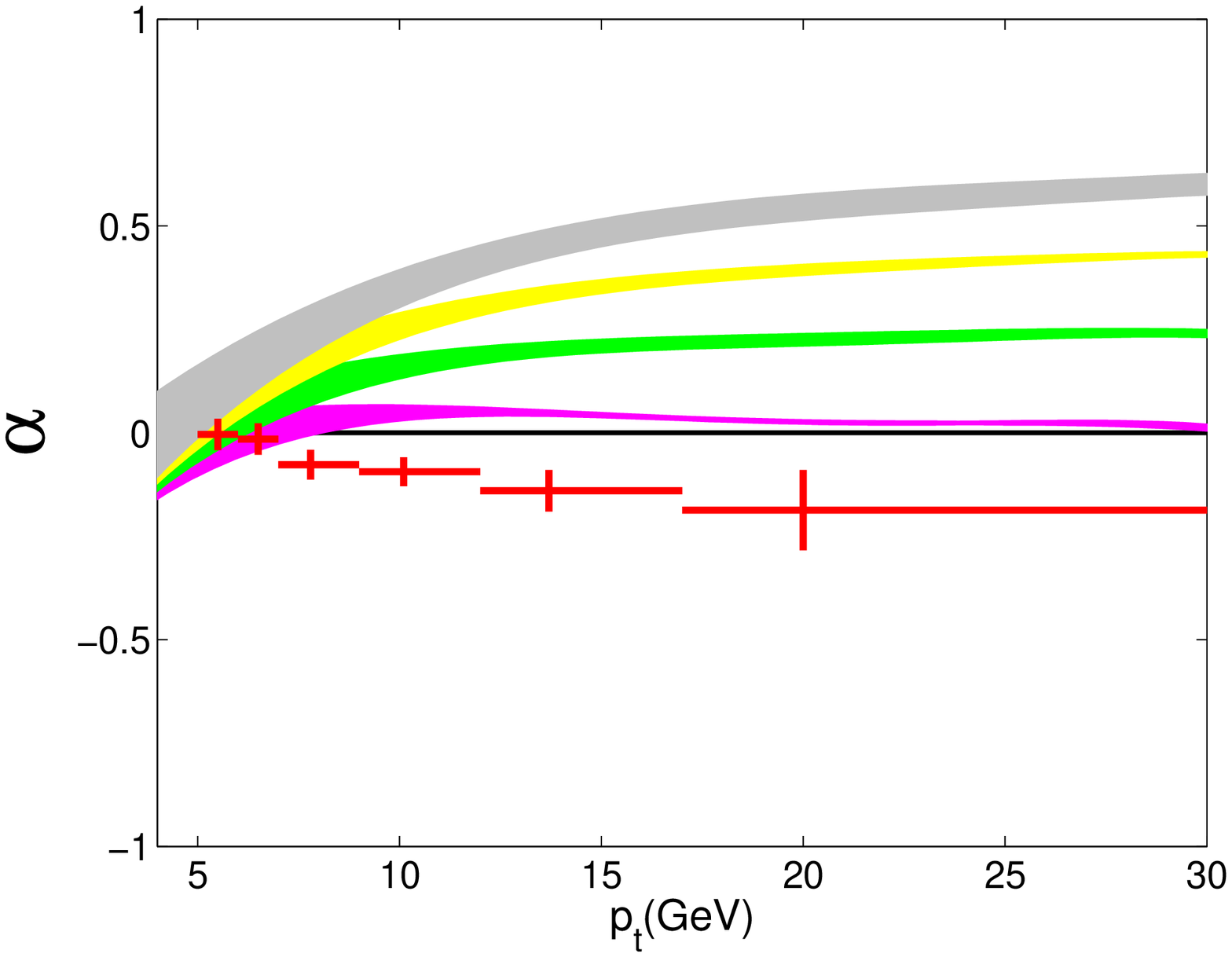}
\caption{$\alpha$ for the prompt $J/\psi$ production. Left diagram
is for $f_1=v f_0$ and the right diagram is for $f_1=f_0$. The upper
shaded band is for $f_0=0$, the higher middle yellow band is for
$f_0=v^4$, the lower middle green band is for $f_0=v^2$ and the
lowest band is for $f_0=1$, where the uncertainties of the matrix
elements are also included. The upper edge of each band is for $x=1$
and the lower edge for $x=0$. The rapidity cut $|y^{J/\psi}|<0.6$ is
adopted. The experimental data on the prompt $J/\psi$ is from
Ref.\cite{expcdfII}. Note the upper shaded band is close to
Braaten's results \cite{braaten}.} \label{ppsi}
\end{figure}

In Fig.(\ref{ppsi}), we present the value of $\alpha$ as a function
of $p_t$ for the prompt $J/\psi$ production, where typical values
for $f_0$ and $f_1$ are adopted and the uncertainties from the
matrix elements are also included. The experimental data on the
prompt $J/\psi$ is from the TEVATRON CDF collaboration
\cite{expcdfII}. It can be found that the results for $f_0=f_1=0$,
i.e. without taking the spin-flipping effects into account, is
consistent with Braaten's results \cite{braaten}. It is clear that
under the proper spin-flip interaction, $\alpha$ can be more closer
to the experimental data than that without these interactions. At
large $p_t$, $\alpha$ is reduced by $\sim50\%$ for $f_0 = v^2$ and
by $\sim80\%$ for $f_0=1$.

\subsection{Numerical results of $\alpha_B$ for indirect $J/\psi$
production through $b\to J/\psi+X$}

\par
\begin{figure}[hbt]
\begin{center}
\includegraphics[width=0.47\textwidth]{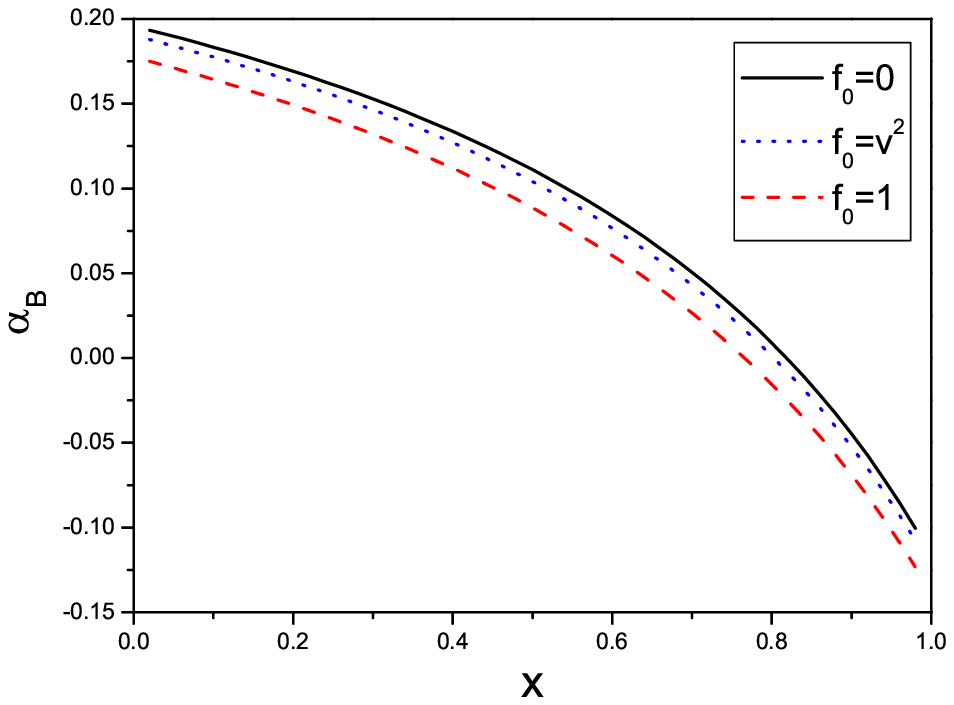}%
\includegraphics[width=0.47\textwidth]{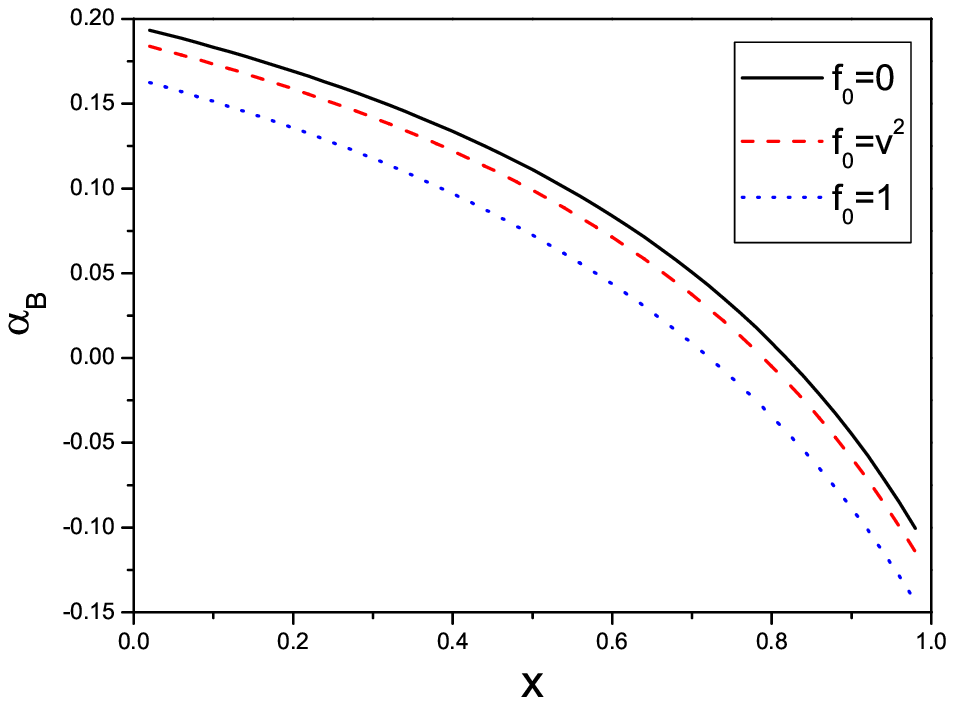}
\end{center}
\caption{The predicted $\alpha_B$ for $b\to J/\psi+X$ as a function
of the unknown parameter $x$. In the left diagram $f_1$ is taken as
$v f_0$, while the right one is for $f_1=f_0$ . } \label{bpialpha}
\end{figure}
\par

By varying the matrix elements within the region of TAB.\ref{tab1},
we calculate the numerical results of $\alpha_B$ versus $x$ for the
indirect $J/\psi$ production through $B\to J/\psi+X$, where
$x=\langle{\cal O}^{J/\psi}_8(^1S_0)\rangle/M^{J/\psi}_{3.4}$. It is
found that the predicted $J/\psi$ polarization parameter $\alpha_B$
depends weaker on $f_0$ and $f_1$ than the case of direct and prompt
$J/\psi$ production. More explicitly under the case of $f_1=v f_0$,
the range of $\alpha_B$ is shifted from $[-0.100,0.193]$ to
$[-0.123,0.175]$ by varying $f_0$ from $0$ to $1$, while under the
case of $f_1=f_0$, the range of $\alpha_B$ is shifted from
$[-0.100,0.193]$ to $[-0.143,0.163]$ by varying $f_0$ from $0$ to
$1$. On the other hand, it dependents heavily on the matrix elements
$\langle{\cal O}^{J/\psi}_8(^1S_0)\rangle$ and $M^{J/\psi}_{3.4}$.
To be consistent with $\alpha_B$ derived in literature, e.g.
$\alpha_B=-0.13\pm0.01$ for $J/\psi$ events with $p_T(J/\psi)>4GeV$
by CDF group\cite{cdfb}, a larger $x$ that approaches 1 should be
taken, as is implicitly adopted by Ref.\cite{btopsi}. Since the
spin-flip effect is weaker in this indirect $J/\psi$ production, a
more precise measurement of it can predict more precise matrix
elements. As a cross check, we have found that without considering
the spin-flip effect and by varying the involved matrix elements
within the same uncertainty region derived by Ref.\cite{btopsi}, we
can obtain the same allowable region for $\alpha_B$, i.e. $\alpha_B
\in [-0.33,0.05]$.

\section{Summary}

It is noted that the $\Upsilon(nS)$ production is somewhat different
from the case of $J/\psi$ and $\psi'$. Ref.\cite{upsilon} shows that
the NLO correction plus the LO results for the color-singlet of
$\Upsilon(nS)$ can explain both the total unpolarized and the
polarized $\Upsilon(nS)$ production cross sections. Then there is no
need to consider the contributions from the color-octet transitions
for $\Upsilon(nS)$ production, or in another words, the color-octet
components give negligible contributions to the $\Upsilon(nS)$
production. For the $J/\psi$ or $\psi'$ production, the NLO
correction plus the LO results for the color-singlet production can
also lead to longitudinal polarized $J/\psi$ or $\psi'$
\cite{wang1}. However, it is well-known that with the color-singlet
contribution only, one can not explain the unpolarized $J/\psi$ or
$\psi'$ production cross section. And by taking the color-octet
$c\bar{c}$ components into consideration, one can well explain the
total unpolarized cross section of $J/\psi$ or $\psi'$
\cite{nrqcdt}, which is regarded as a great triumph of NRQCD. Within
the NRQCD framework, if keeping the spin symmetry for the charm
quark, Refs.\cite{wang1,wang2} show that the $J/\psi$ polarization
puzzle can not be solved even by including the NLO corrections.
Hence to well explain both the unpolarized and longitudinal cross
sections of $J/\psi$ or $\psi'$ is much more involved than the case
of $\Upsilon(nS)$.

We have shown that the spin-flip interaction can have a significant
impact on the transition of a color-octet $^3S_1$ $c\bar{c}$ pair
into $J/\psi$. Such impact can be parameterized by introducing new
parameters in the transition matrix $T$, i.e. $a_{0,1}$ and
$c_{0,1,2}$. If the heavy quark spin symmetry holds, the matrix has
only one parameter $c_0$. The newly introduced parameters $a_{0,1}$
and $c_{1,2}$ are power suppressed in comparison to that of $c_0$ in
principle. However the charm quark is not heavy enough, or $v$ is
not small enough, these new parameters are not small in comparison
with $c_0$, or even they can be at the same size of $c_0$.
Numerically, it is found that these parameters can significantly
reduce the polarization parameter $\alpha$ of $J/\psi$. More
explicitly, we have calculated the direct $J/\psi$ polarization, the
prompt $J/\psi$ polarization and the indirect $J/\psi$ polarization
from the $b$ decays.

Without the spin-flip effect, we return to the same results with
those in literature under the same parameters. While by taking the
spin-flip interaction into consideration, the predicted $\alpha$ is
more close to those measured at TEVATRON. At large $p_t$, $\alpha$
for the prompt $J/\psi$ is reduced by $\sim50\%$ for $f_0 = v^2$ and
by $\sim80\%$ for $f_0=1$. Then such spin-flip interaction as have
been argued by several authors may provide a suitable way to solve
the $J/\psi$ polarization puzzle at TEVATRON. Since the NLO
correction shall provide a large $K$ factor of the total cross
section (ratio of NLO to LO), e.g. $K\sim 2$ for color-singlet
$(c\bar{c})_1[^3S_1]$ and $K\sim 1$ for color-octet
$(c\bar{c})_8[^1S_0]$ and $(c\bar{c})_8[^3S_1]$ \cite{wang1,wang2},
and because the NLO can increase the transverse distributions of
produced $J/\psi$ more than the total distributions, then the value
of $\alpha$ can be further lowed by including NLO results into our
present calculation. Further more, it has been argued that the
production of $J/\psi$ associated with a $c\bar{c}$ quark pair might
also help to dilute the $J/\psi$ polarization \cite{maltoni}. Such
analysis is out of the range of the present paper, which is much
more involved since it involves a NLO calculation of these processes
with spin-flip effects being under consideration and a newly
systematical determination of the color-octet matrix elements. More
over, we have found that the predicted indirect $J/\psi$
polarization parameter $\alpha_B$ depends weaker on the spin-flip
effects than the case of direct and prompt $J/\psi$ production. And
then a more precise measure of the indirect $J/\psi$ polarization
from the $b$ decays can be adopted to predict more precise
color-octet matrix elements, which can inversely improve our
estimations on the direct and prompt $J/\psi$ production.

\vspace{1cm}

\noindent{\bf Acknowledgments}: This work was supported in part by
Natural Science Foundation Project of CQ CSTC under Grant
No.2008BB0298 and Natural Science Foundation of China under Grant
No.10805082, and by the grant from the Chinese Academy of
Engineering Physics under Grant No.2008T0401 and Grant No.2008T0402.
X.G. Wu would also like to thank Prof.J.P. Ma for helpful
discussions on this issue during
his stay in ITP. \\

\section{Expand $\frac{d \hat{\sigma}[{^3S_{1}^{(8)}},gg]}
{{d\hat{t}}}$ in the large $p_t$ limit}

\subsection{Formulae for the production of $J/\psi$
through the channel of $(c\bar{c})_1[^{3}P_2]$}

For the case of $n=(c\bar{c})_1[^{3}P_2]$, the differential
cross-section takes the form:
\begin{equation}\label{totalcs2}
d\sigma_{\lambda}(p\bar{p} \to J/\psi^{\lambda}(n) X) = \sum_{ab}
\int dx_a dx_b f_{a/p} (x_a) f_{b/\bar{p}}(x_b)
d\hat\sigma_{\mu\nu\rho\sigma}[n, ab]
\rho_{|\lambda|}^{\mu\nu\rho\sigma},
\end{equation}
where we have
\begin{eqnarray}\label{totalcs3}
\frac{ d \hat \sigma_{\mu\nu\rho\sigma} [n,ab]}{d t} & = &
A_{ab}[n]g_{\mu\rho}g_{\nu\sigma} + B_{ab}[n]
g_{\mu\rho}k_{1\nu}k_{1\sigma}+ C_{ab}[n]
g_{\mu\rho}k_{2\nu}k_{2\sigma}
+D_{ab}[n]g_{\mu\rho}k_{1\nu}k_{2\sigma}\nonumber\\
&&+E_{ab}[n]k_{1\mu}k_{1\rho}k_{1\nu}k_{1\sigma}
+F_{ab}[n]k_{2\mu}k_{2\rho}k_{2\nu}k_{2\sigma}
+G_{ab}[n]k_{1\mu}k_{1\rho}k_{1\nu}k_{2\sigma}\nonumber\\
&& +H_{ab}[n]k_{2\mu}k_{2\rho}k_{2\nu}k_{1\sigma}
+I_{ab}[n]k_{1\mu}k_{1\rho}k_{2\nu}k_{2\sigma}
+J_{ab}[n]k_{1\mu}k_{2\rho}k_{1\nu}k_{2\sigma} ,
\end{eqnarray}
where
\begin{equation}
\rho^{\mu\nu\rho\sigma}=\sum_{\lambda=-2}^{2}\epsilon^{\mu\nu*}(\lambda)
\epsilon^{\rho\sigma}(\lambda)=\frac{1}{2}(\rho^{\mu\rho}\rho^{\nu\sigma}
+\rho^{\mu\sigma}\rho^{\nu\rho})-\frac{1}{3}\rho^{\mu\nu}
\rho^{\rho\sigma}
\end{equation}
and
\begin{eqnarray}
\rho^{\mu\nu\rho\sigma}_0&=&\epsilon^{\mu\nu*}(0)
\epsilon^{\rho\sigma}(0)=\frac{1}{6}(2\rho_0^{\mu\nu}-\rho^{\mu\nu}_1)
(2\rho_0^{\rho\sigma}-\rho^{\rho\sigma}_1) \\
\rho^{\mu\nu\rho\sigma}_1&=&\sum_{|\lambda|=1}\epsilon^{\mu\nu*}
(\lambda)\epsilon^{\rho\sigma}(\lambda)=\frac{1}{2}(\rho_0^{\mu\rho}
\rho_1^{\nu\sigma}+\rho_0^{\mu\sigma}\rho_1^{\nu\rho}+\rho_0^{\nu\rho}
\rho_1^{\mu\sigma}+\rho_0^{\nu\sigma} \rho_1^{\mu\rho}) \\
\rho^{\mu\nu\rho\sigma}_2&=&\sum_{|\lambda|=2}\epsilon^{\mu\nu*}
(\lambda)\epsilon^{\rho\sigma}(\lambda)=\frac{1}{2}(\rho_1^{\mu\rho}
\rho_1^{\nu\sigma}+\rho_1^{\mu\sigma}\rho_1^{\nu\rho}-\rho_0^{\mu\nu}
\rho_1^{\rho\sigma})
\end{eqnarray}
and
\begin{eqnarray}
\rho^{\mu\nu}&=&\sum_{\lambda=-1}^{1}\epsilon^{\mu*}
(\lambda)\epsilon^{\nu}(\lambda)=-g^{\mu\nu}+\frac{P^{\mu}P^{\nu}}{M^2}\\
\rho^{\mu\nu}_0&=&Z^{\mu}Z^{\nu}\;,\;
\epsilon^{\mu}(0)=Z^{\mu}=\frac{(P\cdot Q/M)P^{\mu}-M Q^{\mu}}
{\sqrt{(P\cdot Q)^2-M^2 Q^2}} \\
\rho^{\mu\nu}_1&=&\sum_{|\lambda|}\epsilon^{\mu*}
(\lambda)\epsilon^{\nu}(\lambda)=\rho^{\mu\nu}-\rho^{\mu\nu}_0 .
\end{eqnarray}
$k_1$ and $k_2$ are the momenta of the initial state partons,
$ab=gg$, $gq$, $g\bar{q}$ and $q\bar{q}$. $P$ and $Q$ are the
momenta of the bound state and the total four-momentum of the
colliding hadrons respectively. All coefficients, $A_{ab}[n]$,
$B_{ab}[n]$, $C_{ab}[n]$ and etc. can be read from
Refs.\cite{AKL,beneke2,kl}.

\subsection{Formulae for the $p_t$-expansion of the dominant
gluon-gluon fusion subprocess}

As for the dominant subprocess: $g(p_1)+g(p_2) \to
J/\psi((^3S_1)_8)(p_3)+g(p_4)$, in the laboratory Frame, we have
\begin{equation}
p_1=\frac{\sqrt{S}}{2}(x_a,0,0,x_a),\;\;
p_2=\frac{\sqrt{S}}{2}(x_b,0,0,-x_b),\;\;
p_3=(M_T\cosh(y),p^x_3,p^y_3,M_T\sinh(y)),\;\;
\end{equation}
where $p_i=(p^0_i,p^x_i,p^y_i,p^z_i)$, $y$ is the rapidity of
$J/\psi$, $M_T^2=M^2+p_t^2$, and we have
\begin{eqnarray}
\hat{s}&=&(p_1+p_2)^2=x_a x_b S,\nonumber\\
\hat{t}&=&(p_1-p_3)^2=M^2-\sqrt{S}M_T x_a
[\cosh(y)-\sinh(y)],\nonumber\\
\hat{u}&=&(p_1-p_4)^2=M^2-\sqrt{S}M_T x_b [\cosh(y)+\sinh(y)],
\end{eqnarray}
with $S$ the square of C.M. energy for the hadronic collider. Using
the above formulae, we obtain the $p_t$-expansion for the dominant
gluon-gluon fusion subprocess in the large $p_t$ limit:
\begin{eqnarray}
\frac{d \hat{\sigma}[(^3S_1)_8,gg]} {{d\hat{t}}}&=&
\frac{3e^{2y}f_1{\pi }^2{\alpha_s }^3}{2M^3S^2{x_a}{x_b}{( {x_a} +
e^{2y}{x_b} ) }^2} -\nonumber\\
&&\frac{1}{p_t}\left[\frac{3e^yf_1{\pi }^2{\alpha_s }^3( M^2{( {x_a}
- e^{2y}{x_b} ) }^2 - S{x_a}{x_b}{( {x_a} + e^{2y}{x_b} ) }^2 )
}{2M^3 S^{\frac{5}{2}}{{x_a}}^2{{x_b}}^2{( {x_a} +
e^{2y}{x_b} ) }^3 }\right] - \nonumber\\
&&\frac{1}{{p_t}^2}\left[\frac{{\pi }^2{\alpha_3 }^3}{36{( 1 +
e^{2y} ) }^2M^3S^3{{x_a}}^3{{x_b}}^3 {( {x_a} + e^{2y}{x_b} )
}^4}\right] \Bigg[ 27( f_2-f_1) S^2{{x_a}}^6{{x_b}}^2
-\nonumber\\
&& 27e^{12y}( f_1 - f_2 ) S^2{{x_a}}^2{{x_b}}^6 + 2e^{10y}{{x_b}}^3(
f_1{x_a} ( 100M^4 - 81M^2S{x_a}{x_b}
+\nonumber\\
&& 27S^2{x_a}( 2{x_a} - {x_b} ) {{x_b}}^2 )+27f_2S{{x_b}}^2 ( -(
M^2{x_b} ) + S{{x_a}}^2( 2{x_a} + {x_b} ) ) ) -\nonumber\\
&& 2e^{2y}{{x_a}}^3 ( f_1{x_b}( -100M^4 + 81M^2S{x_a}{x_b} +
27S^2{{x_a}}^2( {x_a} - 2{x_b} ) {x_b} ) +
\nonumber\\
&& 27f_2S{{x_a}}^2 ( M^2{x_a} - S{{x_b}}^2( {x_a} + 2{x_b} ) ) ) +
e^{8y}{x_a}{{x_b}}^2( f_1 ( -8M^4( 31{x_a} - 50{x_b} )\nonumber\\
&& -324M^2S{x_a}{x_b}( {x_a} + {x_b} ) + 27S^2{x_a}{{x_b}}^2(
10{{x_a}}^2 + 8{x_a}{x_b} - {{x_b}}^2 ) ) +\nonumber\\
&& 27f_2S{{x_b}}^2 ( -4M^2{x_b} + S{x_a} ( 6{{x_a}}^2 +
8{x_a}{x_b} + {{x_b}}^2 ) ) ) +e^{4y}{{x_a}}^2{x_b}\cdot\nonumber\\
&& ( f_1 ( 8M^4( 50{x_a} - 31{x_b} ) -
324M^2S{x_a}{x_b}( {x_a} + {x_b} ) - 27S^2{{x_a}}^2{x_b}( {{x_a}}^2 -\nonumber\\
&&  8{x_a}{x_b} - 10{{x_b}}^2 ) ) + 27f_2S{{x_a}}^2 ( S{x_b} (
{{x_a}}^2 + 8{x_a}{x_b} + 6{{x_b}}^2
)-4M^2{x_a} ) )\nonumber\\
&&  + 2e^{6y}{x_a}{x_b}( -27f_2S{x_a}{x_b} ( M^2(
{{x_a}}^2 + {{x_b}}^2 ) -2S{x_a}{x_b}( {{x_a}}^2 + 3{x_a}{x_b} + {{x_b}}^2 ) ) \nonumber\\
&&  + f_1( -81M^2S{x_a}{x_b} ( {{x_a}}^2 + 4{x_a}{x_b} + {{x_b}}^2 )
+54S^2{{x_a}}^2{{x_b}}^2({{x_a}}^2 + 5{x_a}{x_b} + {{x_b}}^2 )\nonumber\\
&&  + 4M^4( 25{{x_a}}^2 - 62{x_a}{x_b} + 25{{x_b}}^2 ) ) ) \Bigg]+
{\cal O}\left(\frac{1}{p_t^3}\right),
\end{eqnarray}
where $f_1=3a_0 + a_1 + 2c_0$ and $f_2=4c_1+3c_2$. It can be found
that $f_2$ comes into contribution at least at ${\cal O}(1/p_t^2)$
in comparison to $f_1$. While for longitudinal distribution of
$J/\psi$, we only need to make the change: $f_i\to g_i\;\; (i=1,2)$
with $g_1=a_0 + a_1 $, $g_2=2c_0+4c_1 + c_2$, i.e.
\begin{equation}
\frac{d \hat{\sigma}_{L}[(^3S_1)_8,gg]} {{d\hat{t}}}= \frac{d
\hat{\sigma}[{^3S_{1}^{(8)}},gg]} {{d\hat{t}}}(f_1\to g_1; f_2\to
g_2).
\end{equation}
One may also find that for the longitudinal part, $c_0$ comes into
contributions at ${\cal O}(1/p_t^2)$ in comparison to the total
summed results, this is the reason why by taking the spin-symmetry,
i.e. $a_0=a_1=c_1=c_2=0$, the longitudinal contributions should be
neglected at large $p_t$ regions, i.e. the $J/\psi$ is transverse
polarized at large $p_t$ regions.

\end{document}